\documentclass[prb,superscriptaddress,reprint]{revtex4-2}
\usepackage{newtxtext}
\usepackage{newtxmath}
\usepackage{amsmath}
\usepackage{physics}
\usepackage{hyperref}
\hypersetup{colorlinks=true,linkcolor=blue,citecolor=blue,urlcolor=blue}
\usepackage{mathtools}
\usepackage[dvipdfmx]{graphicx}
\usepackage{mhchem}
\usepackage{siunitx}
\usepackage{color}

\newcommand{\He}{ H_\mathrm{e} }
\newcommand{\Hn}{ H_\mathrm{n} }
\newcommand{\Hen}{ H_\mathrm{en} }
\newcommand{\Hsoc}{ H_\mathrm{soc} }

\begin{document}
\title{Chirality-Induced Spin Filtering in Pseudo Jahn-Teller Molecules}
\author{Akihito Kato}
\affiliation{Division of Natural and Environmental Sciences, The Open University of Japan, Chiba 261-8586, Japan}
\author{Hiroshi M. Yamamoto}
\affiliation{Research Center of Integrative Molecular Systems, Institute for Molecular Science, Okazaki, Aichi 444-8585, Japan}
\author{Jun-ichiro Kishine}
\affiliation{Division of Natural and Environmental Sciences, The Open University of Japan, Chiba 261-8586, Japan}
\affiliation{Research Center of Integrative Molecular Systems, Institute for Molecular Science, Okazaki, Aichi 444-8585, Japan}
\date{\today}

\begin{abstract}
  Chirality-induced spin selectivity (CISS) refers to an ability to induce a spin polarization of an electron transmitted through chiral materials.
  An important experimental observation is that incredibly large spin polarization is realized at room temperature
  even for organic molecules that have weak spin-orbit coupling (SOC), although SOC is the only interaction that can manipulate the electrons' spins in the setups.
  Therefore, the mechanism of the CISS needs to be constructed in a way insensitive to or enhancing the magnitude of the SOC strength.
  In this paper, we describe a theoretical study of CISS with a model chiral molecule that belongs to the point group $\mathrm{C}_3$.
  In this molecule, electronic translational and rotational degrees of freedom for an injected electron are coupled to one another via the nuclear vibrational mode with a pseudo Jahn-Teller effect. By properly taking the molecular symmetry as well as the time-reversal symmetry into account and classifying the molecular ground states by their angular- and spin-momentum quantum numbers, we show that the chiral molecule can act as an efficient spin filter. The efficiency of this spin filtering can be nearly independent of the SOC strength in this model, while it well exceeds the spin polarization relying solely on the SOC.
  The nuclear vibrations turned out to have the role of not only mediating the translation-rotation coupling,
  but also enhancing the spin-filtering efficiency.
\end{abstract}
\maketitle

\section{Introduction}
\label{sec:intro}
An object
that cannot be overlapped with its mirror images,
namely,
that lacks the reflection symmetries,
is called chiral.
The chiral symmetry breaking gives rise to abundant functionality~\cite{barron2004Molecular}
such as
chiral magnetism
~\cite{muhlbauer09Skyrmion,togawa2016Symmetry},
chiral phononics
~\cite{zhang2015Chiral,zhu18Observation,kishine2020ChiralityInduced},
chiral photonics
~\cite{kuwata-gonokami2005Giant},
and nonreciprocal conductivity
~\cite{rikken2002Observation,tokura2018Nonreciprocal}.
Recent experimental observations
have confirmed that
the chiral materials exhibit spin-selective phenomena,
including a large spin polarization
of photoelectrons transmitted through
helical molecules
~\cite{gohler2011Spin,mishra2013Spindependent,kettner2015Spin}
and
a large spin dependence
of the current-voltage characteristics
for the tunneling electrons
through helical molecules
~\cite{xie2011Spin,kettner2018ChiralityDependent,suda2019Lightdriven}.
Even enantioseparation
and asymmetric electrochemical reactions
by magnetic electrodes
are reported
~\cite{banerjee-ghosh2018Separation,metzger2020Electron}.
Furthermore,
generation of large spin current
was established
for the inorganic chiral crystals
~\cite{inui2020ChiralityInduced,nabei2020Currentinduced,shiota2021ChiralityInduced}.
These phenomena
that originate from the structural chirality
are collectively termed
the chirality-induced spin selectivity (CISS)
~\cite{naaman2012ChiralInduced,naaman2019Chiral,naaman2020Chiral,evers22Theory},
where the spin is oriented parallel or antiparallel
to the velocity of an injected electron
depending on the materials' handedness.

CISS has been actively studied
and become an interdisciplinary research field
spanning physics, chemistry, and biology.
The reported spin polarization
up to $\SI{60}{\%}$ amounts to an effective magnetic field
of the order of $\SI{100}{\tesla}$
~\cite{naaman2012ChiralInduced},
which is unrealistically strong.
So far, theoretical attempts have been made based on the electron motion in a chiral molecule
in the presence of spin-orbit coupling (SOC)
with dephasing~\cite{guo2012SpinSelective,guo2014Spindependent} or nonunitary~\cite{matityahu2016Spindependent} effect.
These attempts have been made partly because of the necessity of removing the restriction made by the Bardarson theorem~\cite{bardarson08Proof}, in which no spin polarization is allowed due to time-reversal symmetry when two-terminal scattering centers without leakage are considered,
which is, however, not always the case as demonstrated for a two-channel model~\cite{utsumi20Spin}.
Recently, the importance of couplings of electronic or other degrees of freedom has been proposed, including
the Coulomb interactions~\cite{fransson19ChiralityInduced},
the nuclear vibrations and polarons~\cite{du20Vibrationenhanced,zhang20Chiralinduced,fransson20Vibrational},
and substrate-molecule interface effects~\cite{alwan21Spinterface}.
In order to find an alternative explanation,
we would like to understand the fact
that the CISS effect is observed
even in organic molecules
with weak SOC strength
in a way
that does not rely explicitly
on the strength of the SOC.
To do so, we hypothesize that
interplay between nuclear and electronic motions
which are governed by the same symmetry restrictions in chiral materials
has a critical role in the CISS effect,
where the leakage is replaced with a built-in molecular degree of freedom rather than an external outlet.

For our purpose,
we consider a molecule
under the pseudo Jahn-Teller effect~\cite{bersuker2006JahnTeller,bersuker20Jahn},
where the coupling between an electronic translation and rotation is mediated
by the nuclear vibrational degrees of freedom.
The Hamiltonian that includes the electron-nuclear coupling and the SOC satisfies the time-reversal symmetry.
This symmetry confirms that the up-spin and the down-spin states are degenerate forming a Kramers pair.
The member of this degenerate pair can be separated from each other
by its translational direction,
which is made possible by the pseudo Jahn-Teller coupling.
Consequently, the chiral molecule acts as the spin filter by introducing the spin-selective electron injection from an external source, which is distinct from the spin polarizing effect during propagation in the molecule.
The coexistence of the SOC and the (pseudo) Jahn-Teller coupling has an intricate effect on the existence of the orbital degeneracy or the conical intersection~\cite{streltsov20JahnTeller,wang19Hamiltonian}. However, this is essentially irrelevant to the spin-filtering effect proposed in this paper.
We further demonstrate that the performance of this chiral spin filter is determined critically by the spin-selective transmission probability. In this model, angular-momentum (AM) quantum numbers play a crucial role in obtaining quite large efficiency insensitive to the SOC strength.

The remaining part of this paper is organized as follows:
In Sec.~\ref{sec:model}, we introduce the model Hamiltonian of a chiral molecule which belong to the point group $\mathrm{C}_3$. In Sec.~\ref{sec:spin_filter}, we propose the spin-filtering mechanism of the chiral molecule based on the classification of eigenstates by the AM and discuss its condition in Sec.~\ref{sec:spinless}.
In Sec.~\ref{sec:nuclear}, we mention the nuclear vibrational role in the efficient spin filtering.
Section~\ref{sec:conclusion} is devoted to the concluding remarks.

\section{Model}
\label{sec:model}
\subsection{Molecular Basis}
Molecular symmetry shapes a potential field
that es exerted on an electron injected into the molecule
to create a rotational motion from the translational one.
As a result,
the molecular orbital occupied by the injected electron is spanned by both the translational and rotational basis.
This study focuses on
the point group $\mathrm{C}_3$
as the minimal point group that allows chiral structure and has a rotational basis with the one rotating in a direction opposite to the other.
Extending the model to the other point groups $\mathrm{C}_n$ with $n \ge 4$ is straightforward.
Let the $z$ axis be the threefold rotation axis.
We write $\ket{\phi_z}$
for the translational state
that is transformed as $z$,
and $\ket*{\bar{\phi}_z} \coloneqq \Theta \ket{\phi_z}$
for its time reversal
with $\Theta$ being the time-reversal operator.
Inclusion of $\ket*{\bar{\phi}_z}$
allows us to construct
the model Hamiltonian
that satisfies the time-reversal symmetry
in an unambiguous way.
Obviously,
$\expval{\hat{p}_z}{\phi_z} = -\expval*{\hat{p}_z}{\bar{\phi}_z}$
holds,
where $\hat{p}_z$ is the $z$ component
of the electron momentum operator,
and $\expval{\hat{p}_z}{\phi_z} > 0$ is assumed.
Because these bases
belong to the $\mathrm{A}$ irreducible representation of the point group $\mathrm{C}_3$,
they are invariant
under the threefold rotation $C_3$.
The bases for the rotation on the $xy$ plane,
which belong to
the $\mathrm{E}_1$ and $\mathrm{E}_2$
irreducible representations,
are denoted by $\ket{\phi_\pm}$.
These rotational states are transformed
under $C_3$ and $\Theta$ as
\begin{equation}
  C_3 \ket{\phi_\pm}
  = e^{\mp i 2\pi/3} \ket{\phi_\pm}
\end{equation}
and
\begin{equation}
  \Theta \ket{\phi_\pm}
  = \ket{\phi_\mp},
\end{equation}
respectively.

In chiral materials,
translational and rotational motions are coupled with each other~\cite{barron2004Molecular}.
However,
in our model,
the coupling between
the translational and rotational states of an electron
is mediated by the nuclear vibrational modes
belonging to the $\mathrm{e}_1$ and $\mathrm{e}_2$ representation.
Indeed, 
the product representation $\mathrm{e}_i \times \mathrm{E}_i$ for $i=1,2$ includes the $\mathrm{A}$ representation.
Consequently, there arises a finite matrix element $\matrixel*{\mathrm{A}}{\mathrm{e}_i}{\mathrm{E}_i}$ in a symbolic form.
The nuclei in the $\mathrm{e}_{1,2}$ representation also induce the coupling
between the rotational states.
This coupling between degenerate rotational states
causes the spontaneous distortion,
which is known as the Jahn-Teller effect~\cite{bersuker2006JahnTeller,bersuker20Jahn}.
The coupling between nondegenerate ($\mathrm{E}_i$ and $\mathrm{A}$) states can also cause a similar symmetry breaking called the pseudo Jahn-Teller effect~\cite{bersuker2006JahnTeller,bersuker20Jahn}.
The nuclear coordinates in the $\mathrm{e}_{1,2}$ representation
are written as
$Q_\pm = \rho e^{\pm i \varphi}$ with the radius $\rho$ and the angle $\varphi$.
The symmetry operations transform $Q_\pm$
in the following way:
\begin{equation}
  C_3 Q_\pm
  = e^{\pm i 2\pi/3} Q_\pm
\end{equation}
and
\begin{equation}
  \Theta Q_\pm
  = Q_\mp.
\end{equation}

To describe the spin-dependent process,
the electronic spin must be added to the basis.
Let $\ket{\uparrow}$ and $\ket{\downarrow}$
denote the up-spin and down-spin states,
respectively.
They are transformed
under the symmetry operations as
\begin{equation}
  C_3 \ket{\uparrow\!\!/\downarrow}
  = e^{\mp i \pi/3} \ket{\uparrow\!\!/\downarrow}
\end{equation}
and
\begin{equation}
  \Theta \ket{\uparrow\!\!/\downarrow}
  = \pm \ket{\downarrow\!\!/\uparrow}.
\end{equation}

\subsection{Model Hamiltonian}
The Hamiltonian, $H$,
describing an electron
propagating through the chiral molecule
consists of the electronic $\He$,
the nuclear $\Hn$,
and the electron-nuclear coupling $\Hen$ Hamiltonians,
which are all spin independent,
and the SOC Hamiltonian $\Hsoc$,
$H = \He + \Hn + \Hen + \Hsoc$.
The electronic Hamiltonian
is given by
\begin{equation}
  \He
  = \epsilon_\mathrm{tr} ( \ketbra{\phi_z} + \ketbra*{\bar{\phi}_z} )
    + \epsilon_\mathrm{rot} ( \ketbra{\phi_+} + \ketbra{\phi_-} ),
\end{equation}
where we set $\epsilon_\mathrm{rot} \equiv 0$ throughout the paper,
and the nuclear Hamiltonian is described as the two-dimensional harmonic oscillator,
\begin{equation}
  \Hn
  = - \frac{\hbar^2}{2M} \left(
    \frac{1}{\rho} \pdv{\rho}\rho\pdv{\rho}
    + \frac{1}{\rho^2} \pdv[2]{\varphi}
    \right)
    + \frac{M\omega^2}{2} \rho^2,
\end{equation}
where $M$ and $\omega$ are
the nuclear mass and frequency and $\hbar$ is Planck's constant.
For derivation of $\Hen$,
we use the symmetry conditions $[\Hen,C_3]=[\Hen,\Theta]=0$
and take only the first order in $Q_\pm$, 
which becomes (see Ref.~\cite{zeng17General} and Appendix~\ref{sec:derive_Hen} for details)
\begin{align}
  \Hen
  = & V_+ Q_- \ketbra{\phi_z}{\phi_+}
    + V_- Q_+ \ketbra{\phi_z}{\phi_-}
  \notag \\
  & + V_-^\ast Q_- \ketbra*{\bar{\phi}_z}{\phi_+}
    + V_+^\ast Q_+ \ketbra*{\bar{\phi}_z}{\phi_-}
  \notag \\
  & + V_0 Q_- \ketbra{\phi_+}{\phi_-}
    + \mathrm{H.c.},
  \label{eq:H_en}
\end{align}
where $\mathrm{H.c.}$ stands for
the Hermitian conjugate of all the preceding terms.
The first and second lines of the right-hand side of Eq.~\eqref{eq:H_en}
represent the translation-rotation coupling with its strength $V_\pm$,
and the third line represents the rotation-rotation coupling with the coupling strength $V_0$.
Without loss of generality,
these coupling constants can be always set to be real.

Here,
we stress that the chirality imposes
\begin{equation}
  V_+ \ne V_-.
  \label{eq:chirality}
\end{equation}
Actually, the reflection symmetry leads to
$V_+ \equiv V_-$.
We see this by adding a vertical mirror, i.e., considering the case of the point group $\mathrm{C_{3v}}$.
In this case,
the reflection operator $\sigma_\mathrm{v}$
acts on the molecular basis as
$\sigma_\mathrm{v}\ket{\phi_\pm} = \ket{\phi_\mp}$,
$\sigma_\mathrm{v}Q_\pm = Q_\mp$,
and $\sigma_\mathrm{v}\ket{\uparrow\!\!/\downarrow}=\pm\ket{\downarrow\!\!/\uparrow}$.

The only ingredient for the spin-dependent process
in our model
is the SOC.
The derivation of $\Hsoc$ is similar to that of $\Hen$
(see Ref.~\cite{wang19Hamiltonian} and Appendix~\ref{sec:derive_Hsoc} for details):
We use the symmetry conditions
$[ \Hsoc, K ] = 0$ for $K = C_3$ and $\Theta$.
For simplicity, we retain only the zeroth-order terms in $Q_\pm$,
\begin{equation}
  \Hsoc
  = \lambda ( \ketbra{\phi_+} - \ketbra{\phi_-} )
    \otimes \hat{\sigma}_z,
  \label{eq:Hsoc}
\end{equation}
with $\hat{\sigma}_z = \ketbra{\uparrow}{\uparrow} - \ketbra{\downarrow}{\downarrow}$
and the SOC strength $\lambda$.
Equation~\eqref{eq:Hsoc} indicates that spin-momentum locking works in the chiral molecule:
In the cases of $V_+ > V_-$ and $V_+ < V_-$,
the electron propagated with the positive momentum stabilizes the counterclockwise ($\ket{\phi_+}$) and clockwise ($\ket{\phi_-}$) rotations, respectively.
For $\lambda > 0$,
this counterclockwise (clockwise) rotation stabilizes the down-spin (up-spin) state.
Here, we note that
we neglect the spin-flipping components of $\Hsoc$
by disregarding the nuclear dependent contributions.
This treatment is consistent with the experimental observation
~\cite{mishra20LengthDependent} that the spin does not flip in the molecule,
which may be attributed to the large spin energy difference~\cite{inui2020ChiralityInduced}.
It should be pointed out, however,
that such spin-flipping components may be relevant to some of the CISS experiments~\cite{inui2020ChiralityInduced,nabei2020Currentinduced,shiota2021ChiralityInduced}.
This point will be discussed in a later part of this paper.

\subsection{Observables}
To numerically solve the Schr{\"o}dinger equation for $H$,
we need the nuclear eigenstates $\{ \ket{n,m}\}$ to evaluate the matrix elements of $H$~\cite{koizumi94Geometric,requist16Molecular}.
Then,
the product state composed of the chiral molecule and injected electron is written as
$\ket{\Psi} = \sum_{a=z,\bar{z},+,-}\sum_{s=\uparrow,\downarrow}\sum_{n,m} C_{n,m}^{(a,s)} \ket{\phi_a,s}\ket{n,m}$ with $C_{n,m}^{(a,s)}$ being the coupling coefficient
($\phi_{\bar{z}} \equiv \bar{\phi}_z$ is used for convenience).
The electron linear momentum and nuclear AM of $\ket{\Psi}$,
which is analyzed later,
are calculated from
$P = \expval*{\hat{p}_z}{\Psi} = g \expval*{\hat{p}_z}{\phi_z}$,
where the momentum factor
$g = \sum_{s,n,m}[ \abs*{C_{n,m}^{(z,s)}}^2 - \abs*{C_{n,m}^{(\bar{z},s)}}^2]$,
and $L_\mathrm{n} = \expval{(-i\hbar\partial_\varphi)}{\Psi}
= \sum_{a,s,n,m} m \hbar \abs*{C_{n,m}^{(a,s)}}^2$,
respectively, as derived in Appendix~\ref{sec:detail_numerical}.
Hereinafter,
the parameter values for the energy and coupling constants are given
in units of $\hbar\omega$ and $(M\omega/\hbar)^{1/2}$, respectively.

\section{Spin Selective Filtering Effect}
\label{sec:spin_filter}
\subsection{Chirality-Induced Spin Filtering}
\begin{figure}[t]
  \centering
  \includegraphics[width=\linewidth]{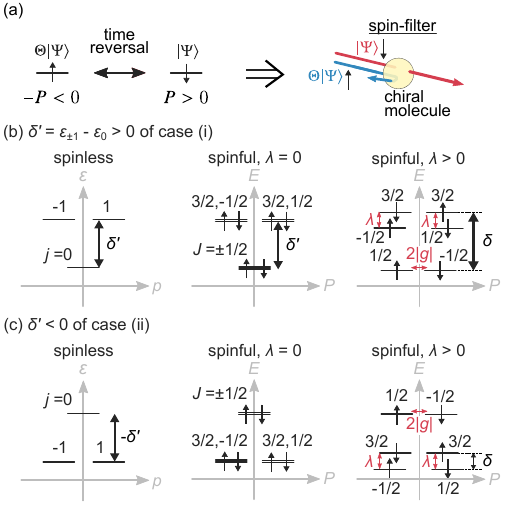}
  \caption{(a) Schematic picture of the spin filer of chiral molecules.
  (b) and (c) Energy-momentum diagrams of spinless, spinful with $\lambda=0$,
  and spinful with $\lambda>0$ systems for (b) $\delta' = \epsilon_{\pm1}-\epsilon_0 > 0$ of case~(i)
  and (c) $\delta' < 0$ of case~(ii).
  Up arrows and down arrows represent the up spin and down spin, respectively.}
  \label{fig1}
\end{figure}

The absence of the spin-flipping components
in $\Hsoc$
indicates that the injected electron passes through the molecule
without changing its spin.
This enables us to propose that
the chiral molecule can act as the spin filter:
Suppose that
one of the ground states, $\ket{\Psi}$,
has positive momentum, $P = \expval*{\hat{p}_z}{\Psi} > 0$, and a down spin.
Due to the time-reversal invariance of $H$,
$\Theta\ket{\Psi}$ is degenerate with $\ket{\Psi}$ but has negative momentum and an up spin.
Thus these states form a Kramers pair.
In injecting a down-spin electron with $P > 0$ into the molecule,
the electron propagates in accordance with $\ket{\Psi}$,
the lowest state among eigenstates with the down spin,
to succeed in passing through the molecule
because of the positivity of the momentum.
In contrast,
the injected up-spin electron with positive $P$ propagates
in accordance with $\Theta\ket{\Psi}$ and not with $\ket{\Psi}$
due to the mismatch of the spin.
However,
because the momentum of $\Theta\ket{\Psi}$ is negative,
the up-spin electron cannot pass through the molecule
[see Fig.~\ref{fig1}(a) for a schematic picture].
Therefore the spin is selectively filtered
depending on the translational direction
of the lowest eigenstate with the same spin as the injected one.
It is of particular importance that
this mechanism of selective spin filtering
requires the molecule to be chiral.
An achiral molecule with $V_+ = V_-$
has a Hamiltonian that is invariant
under the exchange of $\ket{\phi_z}$ and $\ket*{\bar{\phi}_z}$,
physically corresponding to the $z$-inversion invariance of the system.
This invariance
causes the disappearance of the momentum, $\expval*{\hat{p}_z}{\Psi} \equiv 0$,
which means that
achiral molecules cannot utilize the translational direction
of each eigenstate to filter the electronic spins.
Therefore the spin filtering is the chirality-induced effect.
Note that this filtering effect efficiently works
only when the lowest state for the down spin
propagating in the positive translational direction
is energetically well separated from that for the up spin.
We call their energy difference
the spin barrier difference (SBD), denoted by $\delta$,
and explore the condition
that the SBD becomes large enough to explain CISS not relying on the SOC strength $\lambda$.

In what follows,
we consider the chiral case only,
and without loss of generality,
$V_+ > V_- > 0$ is assumed.

\subsection{Classification of Eigenstates by AM Quantum Number}
\label{sec:AM}
The fact that $C_3$ commutes with $H$
allows us to classify all the energy eigenstates
by their AM quantum number $J$.
We write $\ket{\Psi_J}$
for the energy eigenstate with the AM quantum number $J$,
which satisfies the eigenvalue equations
$H\ket{\Psi_J} = E_J\ket{\Psi_J}$
and $C_3\ket{\Psi_J} = e^{-i2\pi J/3}\ket{\Psi_J}$
with $J = 0, \pm 1$ for the spinless systems
and $J = \pm 1/2, 3/2$ for the spinful systems.
Note that
$J$ is determined up to the integer multiple of $3$,
and thus, $J=-3/2$ is equivalent to $J=3/2$.
From the time-reversal invariance,
$\ket{\Psi_J}$ and $\ket{\Psi_{-J}}$ are degenerate,
but with the opposite momentum and spin from each other,
immediately leading to $\expval*{\hat{p}_z}{\Psi_0} = 0$.

Now, we consider the effect of small but finite $\lambda$ on the eigenstates.
For distinction,
we use a capital $J$ and small letters $j$
for the quantum number in the spinful and spinless systems,
respectively,
and corresponding state, energy, and momentum are denoted
by $\ket*{\Psi_J}$, $E_J$, and $P_J$
for the spinful systems
and $\ket*{\psi_j}$, $\epsilon_j$, and $p_j$
for the spinless systems,
respectively.
Two quantum numbers are related by $J = j + 1/2$ for an up spin
and $J = j - 1/2$ for a down spin.
In the cases of $\lambda = 0$,
the spinful states are given by
$\ket*{\Psi_{j+1/2}} \equiv \ket*{\psi_j}\ket{\uparrow}$
or $\ket*{\Psi_{j-1/2}} \equiv \ket*{\psi_j}\ket{\downarrow}$,
where the spatial part of the states is exactly the spinless state $\ket*{\psi_j}$,
indicating that $E_{j\pm1/2} \equiv \epsilon_j$ and $P_{j\pm1/2} \equiv p_j$.
In contrast,
in the cases of $\lambda > 0$,
where the spatial part of $\ket*{\Psi_{j\pm1/2}}$ is modified from $\ket*{\psi_j}$,
these states have different energy and momentum;
for $j=0$, the states with $J = \pm 1/2$ are degenerate
(interchanged by the time-reversal operation)
and $P_{\pm1/2} \ne 0$
due to finite $\lambda$,
as seen in Fig.~\ref{fig2}(a), where the momentum factor $g$ scales linearly with $\lambda$.
The spin direction of the state is governed by the spin-momentum locking.
For $\lambda > 0$ and $V_+ > V_-$,
the momentum of the down spin,
namely, the state with $J = -1/2$, is positive.
For $j=1$ with $p_1 = -p_{-1} > 0$ in the case of $V_+ > V_-$,
the spin-momentum locking stabilizes the down-spin state,
namely, $E_{3/2} > E_{1/2}$.
The origin of the energy splitting in these states is the SOC.

\subsection{SBD}
\label{sec:SBD}
\begin{figure}[t]
  \centering
  \includegraphics[width=\linewidth]{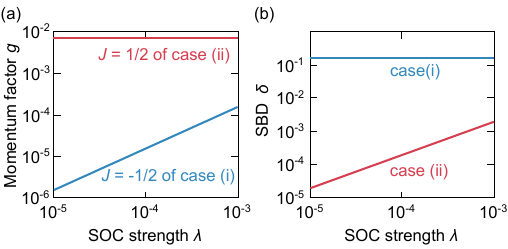}
  \caption{(a) The momentum factor $g$ of the ground state for cases~(i) and (ii) is plotted as a function of the SOC strength $\lambda$.
  Momentum scales linearly for case~(i), whereas it is nearly independent of $\lambda$ for case~(ii).
  (b) The SBDs $\delta \equiv E_{3/2}-E_{-1/2}$ for case~(i) and $\delta \equiv E_{3/2}-E_{1/2}$ for case~(ii) are plotted as a function of $\lambda$.}
  \label{fig2}
\end{figure}

Based on the above argument,
we examine the SBD in the cases of
(i) $\delta' \coloneqq \epsilon_1-\epsilon_0 > 0$
and (ii) $\delta' < 0$.
In case~(i) [see Fig.~\ref{fig1}(b) for a schematic picture],
the chiral molecule filters the spin
with the ground states $\ket*{\Psi_{0\pm1/2}}$.
The SBD is, then,
given by $\delta \equiv E_{3/2} - E_{-1/2} = \delta' + O(\lambda)$.
Hence,
under the condition that $\delta'$ is much larger than the contribution from the SOC,
the SBD is approximated as $\delta \approx \delta'$ and can be nearly independent of $\lambda$, as numerically shown in Fig.~\ref{fig2}(b).
This condition will be discussed later.
We propose that the efficient spin-filtering effect exactly corresponds to this situation, where the effect is insensitive to the magnitude of $\lambda$.
On the other hand,
in case~(ii) [see Fig.~\ref{fig1}(c) for a schematic picture],
the spin filtering occurs
in the ground states $\ket*{\Psi_{1/2=1-1/2}}$
and $\ket*{\Psi_{-1/2=-1+1/2}}$.
The SBD is then identical to $\delta \equiv E_{3/2} - E_{1/2} = O(\lambda)$,
which is the energy difference caused by the SOC
and thus scales linearly with $\lambda$
as presented in Fig.~\ref{fig2}(b).
Therefore the resultant spin-filter efficiency is expected to be small.

\section{Ground-State of Spinless Systems}
\label{sec:spinless}
We justify the conditions under which case~(i) [Fig.~\ref{fig1}(b)] holds, namely, $\ket*{\psi_0}$ becomes the ground state for the spinless systems.
We start this by presenting two limits in which we can analytically identify the AM quantum number of the ground state.
The first is the limit of $V_0 \equiv 0$, where the coupling between the rotational states disappears and only the translation-rotation coupling exists.
In this limit, $\hat{l}_1 = -i\hbar\partial_\varphi - \hbar( \ketbra{\phi_+} - \ketbra{\phi_-} )$ is conserved with the integer eigenvalue $l_1/\hbar \in \mathbb{Z}$.
Among them, the ground state is the state with $l_1 = 0$ that is simultaneously the eigenstate of $C_3$ with $j=0$ (the proof of this statement is given in Appendix~\ref{sec:detal_limit}).
Hence the ground state is given by $\ket*{\psi_0}$ in this limit.
The second limit is the limit of $V_\pm \to 0$,
where the translation and rotation are decoupled,
which means that the injected electron directly passes through or is simply circulating in the molecule.
In this limit,
$\hat{l}_2 = -i\hbar\partial_\varphi + (\hbar/2) ( \ketbra{\phi_+} - \ketbra{\phi_-} )$ is conserved with the half-odd-integer eigenvalue $l_2 /\hbar = \pm (2n-1)/2$ with $n \in \mathbb{N}$.
The ground states are given by the states with $l_2 /\hbar = \pm1/2$ that are simultaneously the eigenstates of $C_3$ with $j=\pm1$.
Therefore, in this limit,
$\ket*{\psi_{\pm1}}$ is the ground state.

\begin{figure}[t]
  \centering
  \includegraphics[width=\linewidth]{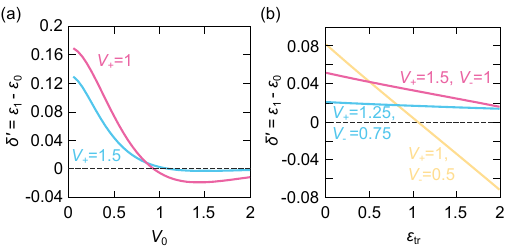}
  \caption{(a) The energy differences of the spinless system $\epsilon_1-\epsilon_0$ are plotted
  as a function of $V_0$ for $V_+ = 1$ and $V_+=1.5$ with fixed $V_- = 0.5$ and $\Delta=0.1$.
  (b) The energy differences of the spinless system $\epsilon_1-\epsilon_0$ are plotted
  as a function of $\epsilon_\mathrm{tr}$ for various values of $V_\pm$ with fixed $V_0 = 0.5$.
  }
  \label{fig3}
\end{figure}

General situations are located in between these two limits
and the ground states are interchanged depending on the parameter values of $\epsilon_\mathrm{tr}$, $V_\pm$, and $V_0$.
Figure~\ref{fig3}(a) presents the energy difference $\delta'$
as a function of the rotational coupling constant $V_0$.
As $V_0$ is decreased below $0.5$,
in which the situation approaches the first limit,
$\delta'$ increases, thus strongly stabilizing the ground state with $j=0$.
Decreasing $\epsilon_\mathrm{tr}$ and/or increasing $V_\pm$
also make  the system approach the first limit,
in which the translation and rotation are strongly coupled.
Therefore the ground state is again given by $\ket{\psi_0}$,
which is consistent with the numerical result presented in Fig.~\ref{fig3}(b).

\section{Role of nuclear AM}
\label{sec:nuclear}
\begin{figure}[t]
  \centering
  \includegraphics[width=\linewidth]{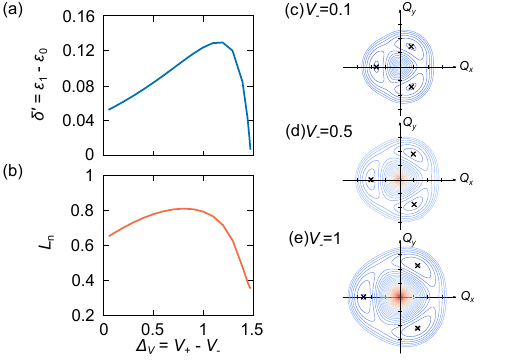}
  \caption{(a) Energy difference $\delta' = \epsilon_1-\epsilon_0$ and (b) nuclear AM $L_\mathrm{n}$ of $\ket{\psi_1}$ are plotted as a function of $\Delta_V = V_+-V_-$ with fixed $V_+$.
  Other parameters are set to $\epsilon_\mathrm{tr}=0.1$ and $V_+ = 1.5$.
  (c)--(e) Ground state Born-Oppenheimer potential surfaces are depicted for $V_- = 0.1$ (c), $0.5$ (d), and $1$ (e).
  Crosses marks the location of the minimal points in each potential surface,
  which are farther from the origin as $V_-$ increases.}
  \label{fig4}
\end{figure}

Finally,
we mention that the energy difference $\delta'$ is correlated with that of the nuclear AM.
Figures~\ref{fig4}(a) and \ref{fig4}(b) present
the energy difference $\delta' = \epsilon_1-\epsilon_0$
and the nuclear AM, $L_\mathrm{n} = \expval*{(-i\hbar\partial_\varphi)}{\psi_1}$,
respectively,
as a function of $\Delta_V \coloneqq V_+ - V_-$ with fixed $V_+$.
The behavior in Figs.~\ref{fig4}(a) and \ref{fig4}(b) can be accounted for by an interplay between the nuclear rotational energy and the pseudo Jahn-Teller effect.
In the $\Delta_V < 0.5$ region in Fig.~\ref{fig4}(a),
as $\Delta_V$ increases,
this imbalance increases the nuclear AM and further differentiates $\ket*{\psi_{\pm1}}$ from $\ket*{\psi_0}$,
thus increasing $\delta'$.
However,
increasing $\Delta_V$ by decreasing $V_-$ also diminishes the pseudo Jahn-Teller effect,
making the nuclear stable configuration close to the high-symmetry point, $\rho \equiv 0$,
and hence reducing the nuclear average radius [see Fig.~\ref{fig4}(c)--(e)],
which leads to the decrease in the nuclear AM and, thus, the nuclear rotational energy.
The resultant energy difference decreases
as presented in the large-$\Delta_V$ region ($\Delta_V > 1.2$) in Fig.~\ref{fig4}(a).
Therefore the nuclear vibrations have the role of not only mediating the translation-rotation coupling,
but also increasing the energy difference $\delta'$, and consequently enhancing the spin-filtering efficiency.

\section{Concluding Remarks}
\label{sec:conclusion}
In this paper,
we have theoretically explored the mechanism
of the spin-filtering effect of chiral molecules.
For that purpose,
we have constructed a model of the point-group $\mathrm{C}_3$
that consists of the electronic translational and rotational states,
which are coupled with each other
by the pseudo Jahn-Teller effect,
or being mediated by the nuclear vibrational degrees of freedom.
The time-reversal symmetry of the Hamiltonian
makes it possible to possess,
in combination with SOC,
spin-momentum locking to filter the injected spin
depending on the translational direction.
Only when the injected spin state matches the lowest eigenstate of the same translational direction
can the electron selectively pass through the molecule with the ground-state energy.
This also enables us to classify the molecular eigenstates
by their AM quantum number.
When the spinless system has a ground state with $j=0$,
the SBD that determines the spin-filter efficiency
is nearly independent of the SOC strength,
and this makes it possible for the chiral molecule to act as an efficient spin filter
protected by the energy for AM quantization.
Here we stress that $\delta'$ can be estimated at around $\SI{50}{\meV}$,
which is about twice as large as room-temperature energy and is therefore enough to explain the $\SI{60}{\%}$ spin polarization in the photoelectron spectroscopy.
In the above estimate, as an example, we employed $\hbar\omega = \SI{0.34}{\eV}$ ($\sim\SI{2740}{\cm^{-1}}$) in Ref.~\cite{allen05Quantum} and a maximum reduced $\delta'$ value of $0.16$ in Fig.~\ref{fig3}~(a).
The ground state with $j=0$ for the spinless systems
is obtained when the translation-rotation coupling constant $V_\pm$ is sufficiently large compared with the rotational one, $V_0$.
In our future work,
the relation between the model parameters and the molecular structure should be clarified together with the direct estimation of the spin-filtering efficiency for actual chiral molecules.
Additionally,
the mechanism of the spin-filtering
proposed in this paper
is not the specific property of the point group $\mathrm{C}_3$ and needs to be straightforwardly generalized to other point or space groups
to which many chiral and/or gyrotropic materials belong.

In this paper, we have omitted the spin-flipping components from the SOC, which may have a role in diminishing the spin-filtering effect to some extent, because SBD in the present example is large enough to justify such a treatment.
From an experimental point of view, this situation is fitting, for example, for the photoelectron transmission through DNA molecules, where the electron is injected from an external source.
However,
a chirality-induced bulk magnetization of \ce{CrNb3S6}
seems to require mechanisms that include the spin-polarizing effect~\cite{nabei2020Currentinduced}.
Therefore CISS seems to include both spin-filtering and spin-polarizing effects depending on the situation.
In our model,
as given explicitly in Appendix~\ref{sec:derive_Hsoc},
we are also able to allow the nuclear-dependent SOC come into play,
yielding spin-flipping terms
such as
$(\ketbra{\phi_+}-\ketbra{\phi_-}) \otimes Q_+ \hat{\sigma}_+$
and
$(\ketbra{\phi_z}-\ketbra*{\bar{\phi}_z}) \otimes Q_+ \hat{\sigma}_+$ with $\hat{\sigma}_+ = \ketbra{\uparrow}{\downarrow}$ and $\hat{\sigma}_- = \ketbra{\downarrow}{\uparrow}$.
These terms allow AM transfer from nuclear phonon to electron spin, thus polarizing the spin in the chiral molecule while respecting the conservation of AM.
This AM conversion may be further enhanced when the molecule involves conical intersections on the adiabatic potential energy surfaces~\cite{wu2021Electronic,bian2021Modeling}.
We here note that the current-induced magnetization phenomenon, called the Edelstein effect,
is apparently similar to the CISS effect from a global symmetry viewpoint.
This effect is, however, interpreted as being mainly dominated by orbital effects~\cite{furukawa2017Observation,yoda2018orbital},
and the resultant magnetization is much weaker than the huge polarization caused by the CISS effect.
The relevance of the spin-flipping components of the SOC to the spin-polarizing effect will be focused on in future studies.

\begin{acknowledgments}
We thank A. S. Ovchinnikov and I. G. Bostrem for discussions in the early stage of this work.
This work was supported by JSPS KAKENHI Grants No. 19H00891, No. 20J01875, and No. 21H01032.
\end{acknowledgments}

\appendix
\section{Derivation of electron-nuclear coupling Hamiltonian}
\label{sec:derive_Hen}
In this appendix,
we derive the electron-nuclear coupling Hamiltonian $\Hen$
following Ref.~\cite{zeng17General}.
The electron-nuclear coupling Hamiltonian $\Hen$ is expanded in terms of the electronic bases $\{ \phi_z, \phi_{\bar{z}} = \bar{\phi}_z, \phi_+, \phi_- \}$,
\begin{equation}
  \Hen
  = \sum_{a \ne b=z,\bar{z},+,-} H_{ab}(Q_+,Q_-) \ketbra{\phi_a}{\phi_b},
\end{equation}
where $H_{ab} = \matrixel*{\phi_a}{\Hen}{\phi_b} = H_{ba}^\ast$ is the nuclear-dependent matrix element.
First, we impose time-reversal symmetry on $\Hen$.
The time-reversal operation $\Theta$ transforms $\Hen$ as
\begin{equation}
  \Theta\Hen\Theta^{-1}
  = \sum_{a,b} (\Theta H_{ab}) \ketbra{\Theta\phi_a}{\Theta\phi_b}
  = \sum_{a,b} H_{ab}^\ast \ketbra{\Theta\phi_a}{\Theta\phi_b}.
\end{equation}
Hence the invariance condition $\Theta\Hen\Theta^{-1} \equiv \Hen$
ensures that the matrix elements must satisfy
\begin{equation}
  H_{\bar{z}-} = H_{z+}^\ast,
  \quad H_{\bar{z}+} = H_{z-}^\ast.
  \label{eq:Hen_TRS}
\end{equation}
In the same manner, the invariance condition under the threefold rotation, $C_3\Hen C_3^{-1} \equiv \Hen$ is equivalent to
\begin{gather}
  C_3 H_{z+} = e^{-i2\pi/3} H_{z+},
  \quad C_3 H_{z-} = e^{i2\pi/3} H_{z-},
  \\
  \quad C_3 H_{+-} = e^{-i2\pi/3} H_{+-}.
  \label{eq:Hen_C3}
\end{gather}
The explicit form of the matrix elements that satisfy Eq.~\eqref{eq:Hen_C3} is given within the first order of $Q_\pm$ as
$H_{z+} = V_+ Q_-$, $H_{z-} = V_- Q_+$, and $H_{+-} = V_0 Q_-$.
Consequently, the electron-nuclear coupling Hamiltonian is obtained as
\begin{align}
  \Hen
  = & V_+ Q_- \ketbra{\phi_z}{\phi_+}
    + V_- Q_+ \ketbra{\phi_z}{\phi_-}
  \notag \\
  & + V_-^\ast Q_- \ketbra*{\bar{\phi}_z}{\phi_+}
    + V_+^\ast Q_+ \ketbra*{\bar{\phi}_z}{\phi_-}
  \notag \\
  & + V_0 Q_- \ketbra{\phi_+}{\phi_-}
    + \mathrm{H.c.},
\end{align}
where $\mathrm{H.c.}$ stands for the Hermitian conjugate of all the preceding terms.

\section{Derivation of Spin-Orbit Coupling Hamiltonian}
\label{sec:derive_Hsoc}
This appendix serves as the derivation of the spin-orbit coupling (SOC) Hamiltonian
following Ref.~\cite{wang19Hamiltonian}.
We employ the Breit-Pauli SOC Hamiltonian,
which can be effectively reduced to a one-electron Hamiltonian
and is expanded in terms of the electronic and spin bases as
\begin{equation}
  \Hsoc
  = \sum_{a,b=z,\bar{z},+,-}\sum_{s_a,s_b=\uparrow,\downarrow}
    f_{as_a,bs_b} \ketbra{\phi_a,s_a}{\phi_b,s_b}.
\end{equation}
For simplicity,
we neglect the matrix elements between the electronic translational and rotational states,
because the transitions between the nondegenerate states are less probable than those between the degenerate states.
For systems with an odd number of electrons,
the time-reversal operator satisfies $\Theta^2 = -1$,
from which the relation for the matrix element of the spin-orbit coupling Hamiltonian
\begin{equation}
  \matrixel{\phi_a,s_a}{\Hsoc\Theta}{\phi_b,s_b}
  = -\matrixel{\phi_b,s_b}{\Hsoc\Theta}{\phi_a,s_a}
\end{equation}
is derived.
This is equivalent to
\begin{align}
  & f_{z\uparrow,\bar{z}\downarrow}
  = f_{z\downarrow,\bar{z}\uparrow}
  = f_{\bar{z}\uparrow,z\downarrow}
  = f_{\bar{z}\downarrow,z\uparrow}
  \notag \\
  & = f_{+\uparrow,-\downarrow}
  = f_{+\downarrow,-\uparrow}
  = f_{-\uparrow,+\downarrow}
  = f_{-\downarrow,+\uparrow}
  = 0
\end{align}
and
\begin{align}
  & f_{z\uparrow,\bar{z}\uparrow} = f_{z\downarrow,\bar{z}\downarrow}, \quad
  f_{z\uparrow,z\downarrow} = -f_{\bar{z}\uparrow,\bar{z}\downarrow},
  \notag \\
  & f_{z\uparrow,z\uparrow} = f_{\bar{z}\downarrow,\bar{z}\downarrow}, \quad
  f_{z\downarrow,z\downarrow} = f_{\bar{z}\uparrow,\bar{z}\uparrow},
  \notag \\
  & f_{z\downarrow,z\uparrow} = -f_{\bar{z}\downarrow,\bar{z}\uparrow}, \quad
  f_{\bar{z}\uparrow,z\uparrow} = f_{\bar{z}\downarrow,z\downarrow},
  \notag \\
  & f_{+\uparrow,-\uparrow} = f_{+\downarrow,-\downarrow}, \quad
  f_{+\uparrow,+\downarrow} = -f_{-\uparrow,-\downarrow},
  \notag \\
  & f_{+\downarrow,+\downarrow} = f_{-\uparrow,-\uparrow}, \quad
  f_{+\downarrow,+\uparrow} = -f_{-\downarrow,-\uparrow},
  \notag \\
  & f_{-\uparrow,+\uparrow} = f_{-\downarrow,+\downarrow}.
  \notag
\end{align}
Next, the Wigner-Eckart theorem can rewrite the matrix element as
\begin{equation}
  \matrixel{\phi_a\uparrow}{H_\mathrm{soc}}{\phi_b\uparrow}
  = \langle\phi_a\|\Hsoc\|\phi_b\rangle
    ( 1/2, 1, 1/2, 0 \mid 1/2, 1/2 ),
\end{equation}
where $\langle\phi_a\|\Hsoc\|\phi_b\rangle$ is the reduced matrix
and $(j_1,m_1,j_2,m_2\mid j,m)$ is the Clebsch-Gordan coefficient.
This decomposition, in combination with the equality
$( 1/2, 1, 1/2, 0 \mid 1/2, 1/2 ) = -( 1/2, 1, -1/2, 0 \mid 1/2, -1/2 )$
results in the following relation
\begin{equation}
  \matrixel{\phi_a\uparrow}{\Hsoc}{\phi_b\uparrow}
  = -\matrixel{\phi_a\downarrow}{\Hsoc}{\phi_b\downarrow}.
\end{equation}
The above relation can be used to relate among the matrix elements as
\begin{alignat}{4}
  f_{z\uparrow,z\uparrow} &= -f_{z\downarrow,z\downarrow}, \quad &
  f_{z\uparrow,\bar{z}\uparrow} &= -f_{z\downarrow,\bar{z}\downarrow}, \quad &
  f_{\bar{z}\uparrow,\bar{z}\uparrow} &= -f_{\bar{z}\downarrow,\bar{z}\downarrow},
  \notag \\
  f_{+\uparrow,+\uparrow} &= -f_{+\downarrow,+\downarrow}, \quad &
  f_{+\uparrow,-\uparrow} &= -f_{+\downarrow,-\downarrow}, \quad &
  f_{-\uparrow,-\uparrow} &= -f_{-\downarrow,-\downarrow}.
  \notag
\end{alignat}
As a consequence of these relations,
the spin-orbit coupling Hamiltonian can be expressed as
\begin{align}
  \Hsoc
  = & f_0 (\ketbra{\phi_+} - \ketbra{\phi_-}) \otimes \hat{\sigma}_z
  \notag \\
  & + f_1 ( \ketbra{\phi_+} - \ketbra{\phi_-} ) \otimes \hat{\sigma}_+
  \notag \\
  & + f_2 ( \ketbra{\phi_z} - \ketbra*{\bar{\phi}_z}) \otimes \hat{\sigma}_z
  \notag \\
  & + f_3 ( \ketbra{\phi_z} - \ketbra*{\bar{\phi}_z}) \otimes \hat{\sigma}_+
  + \mathrm{H.c.},
\end{align}
with $\hat{\sigma}_z = \ketbra{\uparrow}-\ketbra{\downarrow}$,
$\hat{\sigma}_+ = \ketbra{\uparrow}{\downarrow}$,
and $\hat{\sigma}_- = \ketbra{\downarrow}{\uparrow}$.
We impose the invariance under $C_3$ operation on $\Hsoc$,
which gives
\begin{equation}
  C_3 f_0 = f_0, \quad
  C_3 f_1 = e^{i2\pi/3} f_1, \quad
  C_3 f_2 = f_2, \quad
  C_3 f_3 = e^{i2\pi/3} f_3.
\end{equation}
Although these matrix elements $f_k$ ($k=0, \ldots, 3$) depend on $Q_\pm$, we retain only the zeroth-order, i.e., the nuclear-independent, terms. Then, it is legitimate to discard the noninvariant $f_1$ and $f_3$ terms.
Furthermore, the $f_2$ term has the same physical effect as the $f_0$ term for the coupled translation-rotation systems.
Based on this consideration, it is legitimate to retain the $f_0$ term, and 
the relevant spin-orbit coupling Hamiltonian is given by
\begin{equation}
  \Hsoc
  = \lambda (\ketbra{\phi_+} - \ketbra{\phi_-}) \otimes \hat{\sigma}_z,
\end{equation}
with the real-valued constant $\lambda$.

Here we comment on the $Q_\pm$ dependence of the SOC. If the first-order contribution is incorporated,
we have that $\Hsoc$ is given by
\begin{align}
  \Hsoc
  & = \lambda (\ketbra{\phi_+} - \ketbra{\phi_-}) \otimes \hat{\sigma}_z
  \notag \\
  &\, + ( \ketbra{\phi_+} - \ketbra{\phi_-} ) \otimes
  ( \nu Q_+ \hat{\sigma}_+ + \nu^\ast Q_- \hat{\sigma}_- )
  \notag \\
  &\, + ( \ketbra{\phi_z} - \ketbra*{\bar{\phi}_z}) \otimes
  ( \mu Q_+ \hat{\sigma}_+ + \mu^\ast Q_- \hat{\sigma}_- ),
\end{align}
where $\nu$ and $\mu$ are the complex numbers.
The effects of this term are left for future study.

\section{Numerical Details}
\label{sec:detail_numerical}
In this appendix,
we present the details of how to numerically obtain the coupling coefficient $C_{n,m}^{(a,s)}$
of the wavefunction $\ket{\Psi} = \sum_{a=z,\bar{z},+,-}\sum_{s=\uparrow,\downarrow}\sum_{n,m} C_{n,m}^{(a,s)} \ket{\phi_a,s}\ket{n,m}$ and derive the relation between the coupling coefficients and the observables.
To evaluate the matrix elements of the molecular Hamiltonian $H$,
we employ, as the nuclear basis, the eigenstate of the nuclear Hamiltonian $\Hn$
~\cite{koizumi94Geometric,requist16Molecular}
with the eigenvalue $N\hbar\omega$ with $N = ( n - \abs{m} ) / 2$, which is given
for $n = 0, 1, \ldots$ and $m = -n, -n+2, \ldots, n-2, n$, as
\begin{align}
  \Gamma_{n,m}(\rho,\varphi)
  & \coloneqq \braket{\rho,\varphi}{n,m}
  \notag \\
  & = \sqrt{\frac{N!}{\pi(N+\abs{m})!}} e^{im\varphi} e^{-\rho^2/2} \rho^{\abs{m}} L_N^{\abs{m}}(\rho^2),
\end{align}
where $L_N^m(x)$ is the associated Laguerre polynomial that satisfies the following differential equation:
\begin{equation}
  \left[ x \dv[2]{x} + ( m + 1 - x )\dv{x} + N \right] L_N^m(x) = 0
\end{equation}
The matrix element for $\hat{Q}_+$ is computed as
\begin{align}
  & \matrixel*{n_1,m_1}{\hat{Q}_+}{n_2,m_2}
  \notag \\
  & = \delta_{m_1,m_2+1} \left( \delta_{n_1,n_2} \sqrt{ n_1 + m_1 } - \delta_{n_1,n_2-1} \sqrt{ n_1 + 1 } \right)
\end{align}
for $m_2 \ge 0$ and
\begin{align}
  & \matrixel*{n_1,m_1}{\hat{Q}_+}{n_2,m_2}
  \notag \\
  & = \delta_{m_1,m_2+1} \left( \delta_{n_1,n_2} \sqrt{ n_1 - m_1 + 1 } - \delta_{n_1,n_2+1} \sqrt{ n_1 } \right)
\end{align}
for $m_2 < 0$
by using the recurrence relation $L_N^m = L_N^{m+1} - L_{N-1}^{m+1}$
and the integral formula
\begin{equation}
  \int_0^\infty \dd{x} e^{-x} x^{m} L_M^m(x) L_N^m(x)
  = \frac{(N+m)!}{N!} \delta_{M,N}.
\end{equation}
The matrix element for $\hat{Q}_-$ can be calculated
from that for $\hat{Q}_+$ using the relation
$\matrixel*{n_1,m_2}{\hat{Q}_-}{n_2,m_2}=\matrixel*{n_2,m_2}{\hat{Q}_+}{n_1,m_1}$.

Expanding $H$ with respect to the electronic basis $\{ \ket{\phi_z}, \ket*{\bar{\phi}_z} = \ket*{\phi_{\bar{z}}}, \ket{\phi_+}, \ket{\phi_-} \}$ and the above nuclear basis $\{ \ket{n,m} \}_{n,m}$ gives the molecular eigenstate
\begin{equation}
  \ket{\Psi}
  = \sum_{a=z,\bar{z},+,-} \sum_{s=\uparrow,\downarrow} \sum_{n,m}
    C_{n,m}^{(a,s)} \ket{\phi_a,s} \ket{n,m}
\end{equation}
or
\begin{equation}
  \ket{\Psi(R)}
  = \braket{R}{\Psi}
  = \sum_{a=z,\bar{z},+,-}\sum_{s=\uparrow,\downarrow} C^{(a,s)}(R) \ket{\phi_a,s}
\end{equation}
with $C^{(a,s)}(R) \equiv \sum_{n,m} C_{n,m}^{(a,s)} \Gamma_{n,m}(R)$.
By using the above expansion,
the momentum is expressed as
\begin{align}
  \expval*{\hat{p}_z}{\Psi}
  & = \int_0^{2\pi} \dd{\varphi} \int_0^\infty \dd{\rho} \rho
      \expval*{\hat{p}_z}{\Psi(R)}
  \notag \\
  & = \int_0^{2\pi} \dd{\varphi} \int_0^\infty \dd{\rho} \rho
      \expval*{\hat{p}_z}{\phi_z}
  \notag \\
  & \quad \times \sum_{s=\uparrow,\downarrow} \left[ \abs*{C^{(z,s)}(R)}^2 - \abs*{C^{(\bar{z},s)}(R)}^2 \right]
  \notag \\
  & = \expval*{\hat{p}_z}{\phi_z}
      \sum_{s=\uparrow,\downarrow} \sum_{n,m}
      \left[ \abs*{C_{n,m}^{(z,s)}}^2 - \abs*{C_{n,m}^{(\bar{z},s)}}^2 \right].
\end{align}
In this derivation,
the equality $\expval*{\hat{p}_z}{\bar{\phi}_z} = -\expval*{\hat{p}_z}{\phi_z}$
and the fact that the rotational basis on the $xy$ plane has no $z$ component, $\hat{p}_z \ket*{\phi_\pm} = 0$, are used.
Also, the nuclear angular momentum (AM) can be evaluated as
\begin{align}
  \expval*{\hat{L}_\mathrm{n}}{\Psi}
  & = \int_0^{2\pi} \dd{\varphi} \int_0^\infty \dd{\rho} \rho
    \expval*{(-i\hbar\partial_\varphi)}{\Psi(R)}
  \notag \\
  & = \sum_{a=z,\bar{z},+,-} \sum_{s=\uparrow,\downarrow} \sum_{n,m} m\hbar \abs*{C_{n,m}^{(a,s)}}^2.
\end{align}

\section{Details of Limiting Cases}
\label{sec:detal_limit}
In the first limit of $V_0 \to 0$,
$\hat{l}_1 \coloneqq -i\hbar\partial_\varphi - \hbar ( \ketbra{\phi_+} - \ketbra{\phi_-} )$
is conserved with the integer eigenvalue $l_1/\hbar \in \mathbb{Z}$.
The eigenstate with $l_1/\hbar =0$ is written as
\begin{equation}
  \ket{\Phi_0}
  = a_z(\rho)\ket{\phi_z}
    + a_{\bar{z}}(\rho)\ket*{\bar{\phi}_z}
    + e^{i\varphi}a_+(\rho)\ket{\phi_+}
    + e^{-i\varphi}a_-(\rho)\ket{\phi_-},
\end{equation}
where the coefficients $a_z, a_{\bar{z}}, a_+, a_-$ are obtained by
solving the Schr{\"o}dinger equation.
This state is transformed under the threefold rotation $C_3$ as
\begin{align}
  C_3\ket{\Phi_0}
  & = a_z\ket{\phi_z}
      + a_{\bar{z}}\ket*{\bar{\phi}_z}
      + e^{i(\varphi+2\pi/3)}a_+( e^{-i2\pi/3}\ket{\phi_+} )
  \notag \\
  & \quad + e^{-i(\varphi+2\pi/3)}a_-( e^{i2\pi/3}\ket{\phi_-} )
  \notag \\
  & = a_z\ket{\phi_z}
      + a_{\bar{z}}\ket*{\bar{\phi}_z}
      + e^{i\varphi}a_+\ket{\phi_+}
      + e^{-i\varphi}a_-\ket{\phi_-}
  \notag \\
  & = \ket{\Phi_0}.
\end{align}
Thus $\ket{\Phi_0}$ is also the eigenstate of $C_3$ with the AM quantum number $j=0$.

The second limit is $V_\pm = 0$,
where $\hat{l}_2 \coloneqq -i\hbar\partial_\varphi + (\hbar/2) ( \ketbra{\phi_+} - \ketbra{\phi_-} )$ is conserved with the half-odd-integer eigenvalue $l_2/\hbar = \pm (2n-1)/2$ with $n \in \mathbb{N}$.
Its eigenstates with $l_2/\hbar = 1/2$ and $-1/2$ are expressed as
\begin{equation}
  \ket{\Phi_{1/2}}
  = b_+(\rho)\ket{\phi_+}
    + e^{i\varphi}b_-(\rho)\ket{\phi_-},
\end{equation}
and
\begin{equation}
  \ket{\Phi_{-1/2}}
  = e^{-i\varphi}c_+(\rho)\ket{\phi_+}
    + c_-(\rho)\ket{\phi_-},
\end{equation}
respectively.
The threefold rotation $C_3$ transforms these states as
\begin{align}
  C_3\ket{\Phi_{1/2}}
  & = b_+ (e^{-i2\pi/3}\ket{\phi_+})
      + e^{i(\varphi+2\pi/3)}b_-(e^{i2\pi/3}\ket{\phi_-})
  \notag \\
  & = e^{-i2\pi/3}\ket{\Phi_{1/2}}
\end{align}
and
\begin{align}
  C_3\ket{\Phi_{-1/2}}
  & = e^{-i(\varphi+2\pi/3)} c_+ (e^{-i2\pi/3}\ket{\phi_+})
      + c_-(e^{i2\pi/3}\ket{\phi_-})
  \notag \\
  & = e^{i2\pi/3}\ket{\Phi_{-1/2}}.
\end{align}
Thus, as in the first limit,
$\ket{\Phi_{\pm1/2}}$ is also the eigenstate of $C_3$
with the AM quantum number $j = \pm 1$.

\bibliography{references}

\end{document}